\documentclass[a4paper,fleqn]{cas-sc}

\usepackage[numbers]{natbib}

\usepackage{graphicx}
\usepackage{caption}
\usepackage{subcaption}
\usepackage{listings} % For displaying Alloy code as code
\usepackage{xcolor} %For the colors in the Alloy code
\usepackage{alloy-style} %Need to include the file alloy-style.sty - Downloaded from GitHub
\usepackage{hyperref}
\hypersetup{
    colorlinks=true,
    linkcolor=blue,
    filecolor=magenta,      
    urlcolor=cyan,
}
\def\tsc#1{\csdef{#1}{\textsc{\lowercase{#1}}\xspace}}
\tsc{WGM}
\tsc{QE}
\tsc{EP}
\tsc{PMS}
\tsc{BEC}
\tsc{DE}
\author[1]{Ramesh Narasimman}
\author[2]{Izzat Alsmadi}
\address[1]{Syracuse University}
\address[2]{Texas A\&M, San Antonio}

\begin{document}
\let\WriteBookmarks\relax
\def\floatpagepagefraction{1}
\def\textpagefraction{.001}
%\shorttitle{RBAC for Healthcare}
\shorttitle{ }
\shortauthors{Ramesh Narasimman}

\title [mode = title]{RBAC for Healthcare-Infrastructure and data storage }

\begin{abstract}
Role based Access control (RBAC) is the cornerstone of security for any modern organization. In this report, we defined a health-care access control structure based on RBAC. We used Alloy formal logic modeling tool to model and validate system functions. We modeled system static and dynamic or temporal behaviours. We focused on evaluating properties such as integrity, conformance and progress.
\end{abstract}

\maketitle

\section{Introduction}

Data protection and safeguard are of paramount importance in all industries. In Healthcare, it is also a regulatory and compliance requirement since the data collected includes patient's medical, financial and personal information. Exposure of this data will severely impact the patients including financial losses and potentially impact the brand reputation of the organization. Depending on the nature of the data and the size of the exposure, breaches could lead to criminal charges. In this day and age of computing, all these records are electronic and hence require much more sophisticated means of protection from a wide variety of threats. As of 2017, nearly 9 in 10 (86\%)of office-based physicians had adopted any Electronic Health Records (EHRs), and nearly 4 in 5 (80\%) had adopted a certified EHR. Since 2008, office-based physician adoption of any EHRs has more than doubled, from 42\% to 86\%\cite{articleref1}. This data cannot be locked down completely in order to protect it. Doctors and nurses will need access to this data to perform their jobs. Not having access to the right data at the right time could potentially result in loss of life. Hence securing this data should also focus on making this data accessible to people with the right credentials. This necessitates the need for an access management system. As stated by Haraty and Naous \cite{articleref2}, an access control system should:
\begin{itemize}
    \item Prevent unauthorized modifications
    \item Maintain internal and external consistency
    \item Prevent authorized but improper modifications
\end{itemize}
Role-Based Access Control is a simple but powerful concept that helps with this. Usually referred to as RBAC, Role-Based Access control implements access management through Roles. The major components of RBAC are Resources, Actions and Subject. Each role is assigned a set of permissions that encompasses resource(s) and related action(s) that can be performed on that resource. These roles are then assigned to a subject which will allow the subject to perform these actions. The role essentially defines a logical boundary by explicitly defining the permitted actions and hence prevents a subject from being able to perform any actions that are not assigned. These roles can also be based on several factors including time and state of the system. A subject can hold multiple roles and the combined permissions will be the permissions for the subject. Separation of Duties (SOD) is implemented to prevent conflicting permissions from being assigned to the same subject. SOD ensures subjects cannot inherit conflicting permissions when roles are combined. RBAC roles and SOD definitions are based on the roles and responsibilities defined within an organization and are a direct representation of an organization's security policy. 
\begin{figure}
	\centering
	\includegraphics[scale=.75]{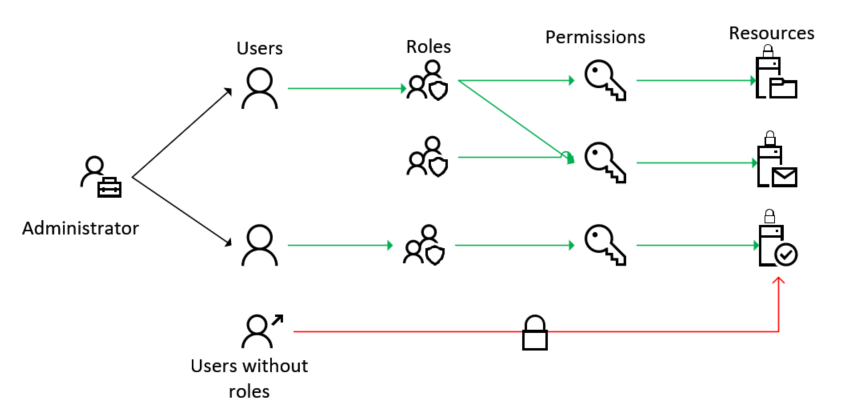}
	\caption{RBAC - high level representation}
	\label{FIG:1}
\end{figure}

Designing a security policy is the first step. For this paper, We used a tool called Security Policy Tool (SPT)\cite{spt} that provides users a way to represent the elements in a tree-like structure. There is a also a command line option, XACML. Security Policy Tool allows us to compose, test, and validate access control policies and to ensure there are no access control leaks when the policies are deployed in a system. With these tests, the access control policies can be effectively analyzed by a policy author to ensure policy works as designed and identify unintended permission assignment. 

The crucial part is having the model tested to ensure it covers all scenarios and produces expected outcome despite user count, multiple roles or other factors. To ensure this, the policy needs to be validated using formal methods. A validated policy is authentic and foolproof. Formal methods are a trusted and efficient way to validate these policies. \emph{Formal methods enable reasoning from logical or mathematical specifications of the behaviors of computing devices or processes; they offer rigorous proofs that all system behaviors meet some desirable property} \cite{chong-nsf-sfm}.  

For this report, We used Alloy to formally specify the policies and validate them. Alloy is a great tool for defining the policy specs, generating a model based on specs and validate the model for correctness. It is a modeling tool based on first order logic combined with relational Algebra. It also includes an analyzer that performs bounded analysis of model specifications and provides a visual representation. Alloy can also evaluate the models and provide counter examples if applicable. Alloy, being based loosely on object oriented language concepts, makes it easy to write code. Its structure includes objects called \emph{Signatures}. It also includes \emph{Facts} that can be used to constrain the models. Additional operations can be performed using \emph{Predicates and Functions}. \emph{Assertions} can be used to validate the constraints placed on the model. Alloy uses SAT solver to analyse the model and establish validity or provide counter examples. 

There are different variations of the Alloy tool. Two are most notable: Sterling and Electrum. They both use Alloy as the base and add additional modules to enhance the functionality. Sterling combines Alloy with web-based visualizations, providing both basic Alloy visualization capabilities and a robust platform for the development of domain specific visualizations of Alloy instances. The Alloy Instances tool provides a styling language and sharing platform that is useful for development of visualizations once a model has been completed, but it lacks in utility during the iterative modeling process.Sterling aims to bridge this gap and further build out the visualization and sharing platforms. Sterling is an Open Source tool based on Alloy as well as \emph{Spark, Gradle Shadow Plugin} and additional JavaScript libraries. 

Using these tools, We have created three basic policies to demonstrate how these tools can be effectively applied in a Healthcare infrastructure. The policies are simple and cover only a finite resources since the scope of this paper is purely academic. However they do incorporate all the necessary attributes and constructs to demonstrate the various aspects of a typical RBAC policy. Also, since RBAC policies closely resemble an organization's security policies, there is no single template that could be created that fits all. 

\section{Literature Review}
We have reviewed literature that are related to Role Based Access Control, Separation of Duties, Temporal-RBAC as well as various formal languages for model analysis. 

Chuck Easttom \cite{secml} proposes a modeling language for cyber security. He discusses the need for a modeling language in cyber security similar to SysML. Cybersecurity can be defined as system of systems. That allows SysML to be used as foundation for the proposed language. With SysML as a base, he proposes a new language that adds additional diagrams that are designed specifically for Cyber security. This new language provides greater flexibility to the security engineers in designing the models and can be used as a generalized security modeling language.

In this next paper, Sommestad, Ekstedt and Holm \cite{Cysemol} discuss about Cyber Security Modeling Language (CySeMoL) and its effectiveness in identifying potential attacks against a system. This language also includes a probabilistic inference engine designed based on compilation of research results on a number of security domains and a wide range of attacks and countermeasures. CySeMoL aims to fill one of the biggest gaps in designing a security system - lack of understanding of the vulnerabilities in different components that interact with the system as well as within the system. CySeMoL is based on logical relations, experimental research and covers a variety of attacks, including software exploits, flooding attacks, abuse of obtained privileges, and social-engineering attacks. It is easy to use and does not require security expertise. A Turing test indicates that the reasonableness and correctness of CySeMoL assessments compare to those of a security professional. 

Johnson, Lagerström and Ekstedt \cite{mal} discuss the need for a reusable, domain independent attack logic that can be utilized to codify attack steps and their dependencies. The article proposes \emph{Meta Attack Language (MAL)} as a potential solution and details the specification and discusses features to help with the challenges in identifying the data to collect, issues in collecting the data and analyze and the article recommends the use of attack simulations. These simulations are based on attack graphs. Since many systems are similar, the attack graphs can be replaced by identifying common factors and import them into a modeling language. This can then be reused across multiple systems without the need for recreating the graphs.

Marrone, et-al discuss the need for vulnerability analysis in critical infrastructure systems such as railway systems. This paper reviews the existing system and provides a specification language CIP-VAM and a model-driven framework, METRIP, based on UML for performing vulnerability analysis and protection modeling functionality. This model takes one station and models the security in that station identifying the gaps and pros.

The next report reviewed \cite{scada} that discusses the need for formal verification of security models for Industrial control systems and how the improvements in IoT is exposing these once-isolated systems to cybersecurity threats. Specific areas of concern are securing the data and the control flow.The proposed model in this paper consists of the control network layer and a PC that connects to the control network for data transfer and management. The PC is the component running the SCADA software that connects externally to the cloud. There are also mobile components that access the data. The paper defines the Security requirements based on the \emph{IEC-62443-3-3} standard and defines behavioral and security properties. The paper also provides the verification as well as the code that is written in the proposed model that includes all the specified requirements.

Chong et. and all \cite{chong-nsf-sfm} review and document the security issues in major verticals including Hardware Architecture, Operating Systems, Distributed Systems and Networks as well as privacy and how formal methods can help resolve these. The report also stresses the importance of formal methods and the benefits it can provide in designing a system that meets the standards, makes interfaces easy, accurate and eliminate potential threats. The report also discusses various problems including ensuring whole-system security, defining correct abstractions, standardizing tool and techniques, ensuring formal methods are supported throughout the life cycle of the product and integrating formal methods with the common industry process in a scalable and compatible way. The workshop provides several recommendations to improve the efficiency and to improve the available talent pool that can help with the increased implementations of formal methods. Those include teaching formal methods early in the education curriculum sooner than they are now and include much more complex problems. Additional recommendations include encouraging community-based development of tools and techniques including conferences, journals as well as repositories. The workshop recommends that focus should be placed on both scientific work and applied engineering to improve cybersecurity through formal methods.

The focus of the next report, a thesis by Nicole Emerentiana van Deursen \cite{hirisk}, is a new method to assess socio-technical information security risks.  The core of this is method is information sharing across healthcare organizations as this could lead to collective learning and problem solving when faced with similar issues.The aim is to reach a collective state of information security that creates trust in the information we collect and share and ensures its validity.This method relies on a central database shared by various organizations. Incidents reported to this database are analyzed, validated by experts and combined with their insights in dealing with similar issues. The results are then transformed into a risk map that contains actionable information and prevention methodologies that can be used by participating healthcare organizations. The map is constantly updated as new incidents are reported.This method differs from its closest competitors due to its indefinite scope and wider system context.

The next report reviewed is also a thesis by Shan Wu \cite{infosec}. This thesis reviews the methods for the development of information security policies at organizations. The author stresses that the goal of information security policies should be to protect the business than focusing on specific data sets. The author then also details the functions each security policy much adhere to including stated punishment and also lists the stakeholders that should be involved in the policy development life cycle.

The next report \cite{trustmodel} is part of the document detailing the \emph{Proceedings of the 10th international Joint Conference On Biomedical Engineering Systems and Technologies} and was authored by \emph{Raja Manzar Abbas, Noel Carroll, Ita Richardson and Sarah Beecham}.This report explores what determines the trustworthiness of healthcare software solutions.It also explores the potential risk of stakeholders trusting the software without realizing the consequences and proposes a model to assess the trustworthiness. While medical devices are subject to strict regulations and constant monitoring by authorities, Healthcare software is not. This leads to questions about the trustworthiness of the applications and the need to evaluate them against a standard. Though there is no proposed model, the report recommends that Security, Efficiency, Safety, Functionality, Reliability, Regulation, Validity and Accuracy be the key factors in the standard to determine the trustworthiness of this software.

The next report deals with Eager formal methods. Funded by the NSA, this report \cite{infoflow} looks at developing a highly abstract model of SELinux access control, map the flow and use model checking to validate the design and establish this as a case study for implementing Eager formal methods. This report also demonstrates how a combination of the formal models and an appropriate algorithm like SELinux provides the necessary framework to improve security in the operating systems and a model that can be replicated easily across any similar systems.

The next report by Haraty and Naous \cite{rbacval} focuses on validating the RBAC aspect of commercial applications using formal specifications. They model the RBAC components using Alloy and check the model consistency. This paper uses an example of a procurement system to demonstrate the implementation of RBAC and the process of modeling and validating using Alloy. 

GEO-RBAC paper summarizes the concept of spatial awareness for RBAC and expands it further to include geographic features/boundaries \cite{spatrbac}. This model not only checks whether the subject has the necessary attributes for role assignment but also looks to ensure the subject is physically present in the location to be able to get the role. This can be enforced using GPS and other location-tracking mechanism. This model also applies the Separation of Duties on to the GEO-RBAC model further strengthening it. The model is designed to be dynamic enough to automatically add or remove roles based on locations.

The problem of multiple organizations existing in the same setup being forced to share access control mechanisms is discussed in \cite{multidomainrbac}. A framework is proposed for these policies to be merged while preserving all permissions and SOD requirements. The framework also proposes methodologies to resolve conflicts as well as a way to identify and remove potentially conflicting permissions while maintaining the original requirements. 

\emph{TrustBAC} describes the challenges in implementing RBAC in open systems as well as the shortcomings of combining RBAC with Credential based Access control \cite{trustrbac}. The authors propose a solution that introduces the trust level in this model. Users who are authenticated using the credentials will be assigned to a trust level based on their behavioural history and these trust level will be mapped to Roles which in turn are mapped to permissions. By introducing the trust levels, the model is much more dynamic and secure.

\section{Policy Definitions}
In this paper, we created three policies that implement basic RBAC concepts. All three policies are based on these objects:
\begin{itemize}
    \item Resource
    \item Role
    \item Action
\end{itemize}
The roles were assigned action(s) for individual resources that are then applied to users. User with the roles will be able to perform the allowed actions. They will not be able to perform anything that is not included in the permissions list. In addition, to ensure compliance and to ensure conflicting permissions are not inherited, separation of duties was also implemented. 

Policy 1 deals with access for three servers - DC1, DC2 and FS1. Traditional roles including System Engineer are defined. Permissions for specific actions on the resources are defined. Policy 2 deals with access for Applications, related servers and database servers. Multiple roles are defined, including the Application Administrator. Permissions are defined based on specific actions on resources. Policy 3 is specifically about security logs and log aggregator devices. 
\begin{table}[width=.9\linewidth,cols=2,pos=ht]
\caption{Policy Resource Summary}\label{polsumtbl}
\begin{tabular*}{\tblwidth}{@{} LLLL@{} }
\toprule
Policy & Resources\\
\midrule
Policy 1 & Domain Controller 1, Domain Controller 2, File Server\\
\midrule
Policy 2 & App Servers, DB servers, Application\\
\midrule
Policy 3 & Servers, Network Devices, Log Aggregator\\
\bottomrule
\end{tabular*}
\end{table}

Detailed policy definitions are provided below:
\subsection{Policy 1: Infrastructure Server Access for Different Roles}
This policy defines access levels for different roles in the company. Access levels defined are:
\begin{itemize}
    \item Admin Access to Servers
    \item RDP Access to Servers
    \item Share access to Servers
\end{itemize}
Permissions are defined for the following roles:
\begin{itemize}
    \item System Engineer
    \item Network Engineer
    \item DBA
    \item IT Manager
    \item End Users
    \item External Users
\end{itemize}
Requirements are defined as follows:
\begin{itemize}
    \item External users are treated as unauthorized and should not have any access to any resources.
    \item IT Manager should not have Admin or RDP access to servers. This is for separation of duties.
    \item Network Engineers and DBA should not have ADMIN or RDP Access to any of the servers.
    \item Only System Engineers should have Admin as well as RDP Access to the Servers.
    \item All users with the exception of external users, should have access to the shares.
\end{itemize}

\subsection{Policy 2: Application and Database Servers}
This policy defines access rules for the application and database servers as well as PHI and De-Identified data in the servers. Access Levels defined are:
\begin{itemize}
    \item ADMIN Access to Application Servers
    \item ADMIN Access to Database Servers
    \item Database ADMIN access
    \item Access to Production Application
    \item Access to Production Data
    \item Access to Test Application
    \item Access to Test/De-Identified Data
\end{itemize}
Permissions are defined for the following roles:
\begin{itemize}
    \item Developer
    \item Tester
    \item Server Administrator
    \item DBA
    \item Application ADMIN Users
    \item Application Users
    \item Other Users
\end{itemize}
Requirements are defined as follows:
\begin{itemize}
    \item Production Servers
    \begin{itemize}
        \item Developers and Testers will not have any access
        \item Server Admins will have full Admin access to both Application and Database Servers
        \item DBA will have Admin access only on the Database servers
        \item DBA will also have Database Access
        \item None of the users will have any access to the servers.
    \end{itemize}
    \item Test Servers
    \begin{itemize}
        \item Developers and Testers will not have any access
        \item Server Admins will have full Admin access to both Application and Database Servers
        \item DBA will have Admin access only on the Database servers
        \item DBA will also have Database Access
        \item None of the users will have any access to the servers.
    \end{itemize}
    \item Production Application
    \begin{itemize}
        \item Developers and Testers will not have any access
        \item Server Admins will not have any access
        \item DBA will not have any access
        \item Application Admin users will only have access to the Admin Screens
        \item Application End users will only have access to Non-Admin screens
    \end{itemize}
    \item Test Application
    \begin{itemize}
        \item Developers and Testers will have full access
        \item Server Admins will not have any access
        \item DBA will not have any access
        \item Application Admin users will only have access to the Admin Screens
        \item Application End users will only have access to Non-Admin screens
    \end{itemize}
    \item Prod PHI/PII Application Data
    \begin{itemize}
        \item Only DBA and Application users will have access.
    \end{itemize}
    \item De-Identified Application Data
    \begin{itemize}
        \item Testers, Developers, DBA, and Application users will have access. 
        \item Server Admins and Other Users will not have access
    \end{itemize}
\end{itemize}

\subsection{Policy 3: Log Access}
This policy defines access levels for Server and Network Security logs. Access levels defined are:
\begin{itemize}
    \item Admin Access
    \item Log Review
    \item Log Access
\end{itemize}

Permissions are defined for the following roles:
\begin{itemize}
    \item Server Engineer
    \item Network Engineer
    \item Security Engineer
\end{itemize}

Requirements are defined as follows:
\begin{itemize}
    \item Server Access is restricted to Server Engineers. Both Server Engineer and Security Engineer will have log review access on servers
    \item Network Device access is restricted to Network Engineers
    \item Both Network Engineer and Security Engineer will have log review access on network devices
    \item Log Aggregator access is restricted to Security Engineer. 
\end{itemize}

\begin{table}[width=.9\linewidth,cols=3,pos=ht]
\caption{Policy Role Summary}\label{polrolesumtbl}
\begin{tabular*}{\tblwidth}{@{} LLLL@{} }
\toprule
Policy & Resources\\
\midrule
Policy 1 & System Admin, Network Engineer, DBA, IT Manager, End Users, External Users\\
\midrule
Policy 2 & Developer, Tester, Server Admins, DBA, App ADMIN, End Users, Other Users\\
\midrule
Policy 3 & Server Engineer, Network Engineer, Security Engineer\\
\bottomrule
\end{tabular*}
\end{table}

\begin{table}[width=.9\linewidth,cols=3,pos=ht]
\caption{Policy Actions Summary}\label{polactsumtbl}
\begin{tabular*}{\tblwidth}{@{} LLLL@{} }
\toprule
Policy & Actions\\
\midrule
Policy 1 & Admin Access, Share Access, RDP Access\\
\midrule
Policy 2 & Admin Access, DB Access, App Access, AppAdmin Access, Data Access\\
\midrule
Policy 3 & Admin Access, Log Review\\
\bottomrule
\end{tabular*}
\end{table}

\section{Security Policy Definition - Using Security Policy Tool}
Based on the policy definitions, policy structure was created using the Security Policy Tool. Subject, Resources and Actions were defined and combined to create Access control policies using \emph{Deny-overrides} Algorithm and \emph{Deny Biased} Policy Enforcement Algorithm. Verification cases were defined to check for the validity of the policies. The tool can be downloaded from \url{https://www.secruitypolicytool.com}.   
\subsection{Policy 1: Infrastructure Server Access for Different Roles}
\begin{figure}[ht]
	\centering
    \begin{subfigure}[b]{0.2\textwidth}
    	\includegraphics[width=\textwidth]{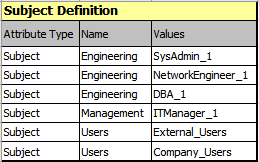}
    	\caption{Subject Definition}
    	\label{policy1_subject}
	 \end{subfigure}
	 \hfill
	 \begin{subfigure}[b]{0.2\textwidth}
    	\includegraphics[width=\textwidth]{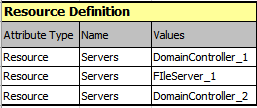}
    	\caption{Resource Definition}
    	\label{policy1_resource}
	 \end{subfigure}
	 \hfill
	 \begin{subfigure}[b]{0.2\textwidth}
    	\includegraphics[width=\textwidth]{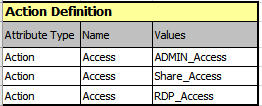}
    	\caption{Resource Definition}
    	\label{policy1_action}
	 \end{subfigure}
	 \caption{Policy 1 - Definitions}
	 \label{policy1_def}
\end{figure}

As defined, this policy dictates and controls access to servers for various roles in the company and specifies separation of duties as well. Definitions are as shown in Figure \ref{policy1_def}. Based on these definitions, access control policies were defined. 
\subsection{Policy 2: Application and Database Servers}
\begin{figure}[ht]
	\centering
    \begin{subfigure}[b]{0.2\textwidth}
    	\includegraphics[width=\textwidth]{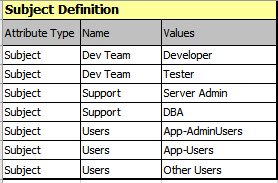}
    	\caption{Subject Definition}
    	\label{policy2_subject}
	 \end{subfigure}
	 \hfill
	 \begin{subfigure}[b]{0.2\textwidth}
    	\includegraphics[width=\textwidth]{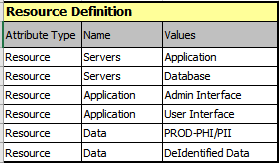}
    	\caption{Resource Definition}
    	\label{policy2_resource}
	 \end{subfigure}
	 \hfill
	 \begin{subfigure}[b]{0.2\textwidth}
    	\includegraphics[width=\textwidth]{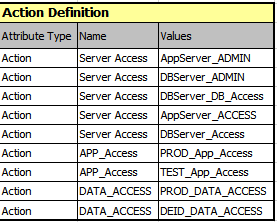}
    	\caption{Resource Definition}
    	\label{policy2_action}
	 \end{subfigure}
	 \caption{Policy 2 - Definitions}
	 \label{policy2_def}
\end{figure}

This policy defines access rules for the application and database servers as well as Protected Health Information (PHI) and De-Identified data in the servers. Definitions are as shown in Figure \ref{policy2_def}. Based on these definitions, access control policies were defined. 
\subsection{Policy 3: Log Access}
\begin{figure}[ht]
	\centering
    \begin{subfigure}[b]{0.2\textwidth}
    	\includegraphics[width=\textwidth]{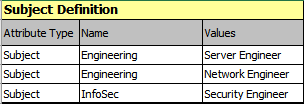}
    	\caption{Subject Definition}
    	\label{policy3_subject}
	 \end{subfigure}
	 \hfill
	 \begin{subfigure}[b]{0.2\textwidth}
    	\includegraphics[width=\textwidth]{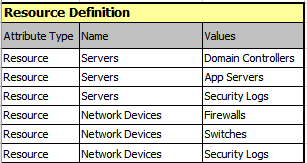}
    	\caption{Resource Definition}
    	\label{policy3_resource}
	 \end{subfigure}
	 \hfill
	 \begin{subfigure}[b]{0.2\textwidth}
    	\includegraphics[width=\textwidth]{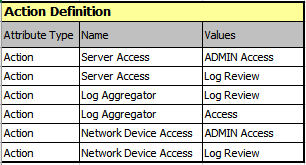}
    	\caption{Resource Definition}
    	\label{policy3_action}
	 \end{subfigure}
	 \caption{Policy 3 - Definitions}
	 \label{policy3_def}
\end{figure}

This policy defines access levels for Server and Network Security logs and centralized Log Aggregator. Definitions are as shown in Figure \ref{policy3_def}. Based on these definitions, access control policies were defined as shown. 

\newpage
\section{Formal Specification of the models - Alloy}
With the policies defined in the security policy tool, the next step is to formally specify the models using Alloy. Each policy is modeled separately. The objects defined in the policy are created as Sigs in Alloy and the ACLs are implemented through predicates. Facts are also used to enforce constraints on the model. 
\subsection{Policy 1 - Alloy specification}
Based on the policy definition, below are the Sigs defined for Policy 1.
\begin{lstlisting}[language=alloy]
abstract sig Role{} //defines the roles
abstract sig Resource{} //defines the Resources
abstract sig Action{} //defines the Actions or Access Levels
\end{lstlisting}
In addition to the Sigs, we added time and state for adding the temporal aspects to the model. The state will determine permissions based on the current status of the system.

\begin{lstlisting}[language=alloy]
sig Time{} //for time definitions
sig RState{}
\end{lstlisting}
With the definitions, several facts were added to model the policy. These facts enforce that every user must have one predefined role and the External Users role cannot be combined with any other roles. To demonstrate the role assignments, we created users that are derived from the user Sig and will inherit the role(s). The roles are listed in Table \ref{pol1usersumtbl}.

\begin{lstlisting}[language=alloy]
fact noOtherRolesforEU {
	no u:User | ExtUsers in u.urole and some (SysAdmin + DBA + NetworkEngineer + ITManager + EndUsers) & u.urole
}
//Fact 2 - All Users must have one of preDefined role
fact {
	all u:User | u.urole in Role
}
\end{lstlisting}
Assertions are added to validate these facts. 
\begin{lstlisting}[language=alloy]
assert rolesNeededforAll {
	all u:User | u.urole in Role
}
assert ExtUserAccess{
	no u:User | ExtUsers in u.urole and some (SysAdmin + DBA + NetworkEngineer + ITManager + EndUsers) & u.urole
}
\end{lstlisting}
With the basic blocks defined, we added predicates to build the permissions. A predicate was defined for each role to define the permission set for that role that details all the actions that the role can perform on the resources. Below is one of the predicates.
\begin{lstlisting}[language=alloy]
pred saPerms[rs:RState]
{
	all usa:SysAdmin,fs:FS | usa.permissions =rs+ fs +AdminAccess and usa.permissions =  rs + fs + ShareAccess and usa.permissions = rs + fs + RDPAccess

	all usa:SysAdmin,dc1:DC1 | usa.permissions = rs + dc1 + AdminAccess + rs and usa.permissions =  rs + dc1 + ShareAccess + rs and usa.permissions = rs + dc1 + RDPAccess + rs

	all usa:SysAdmin,dc2:DC2 | usa.permissions = rs + dc2 + AdminAccess and usa.permissions =  rs + dc2 + ShareAccess and usa.permissions =  rs + dc2 + RDPAccess
}
\end{lstlisting}

\begin{figure}
	\centering
	\includegraphics[scale=.40]{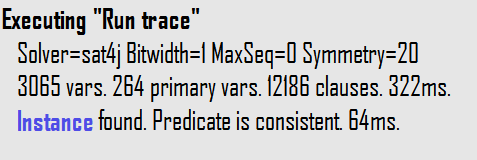}
	\caption{Policy 1 - Alloy Output}
	\label{pol1_alloy_output}
\end{figure}

\begin{figure}
	\centering
	\includegraphics[scale=.40]{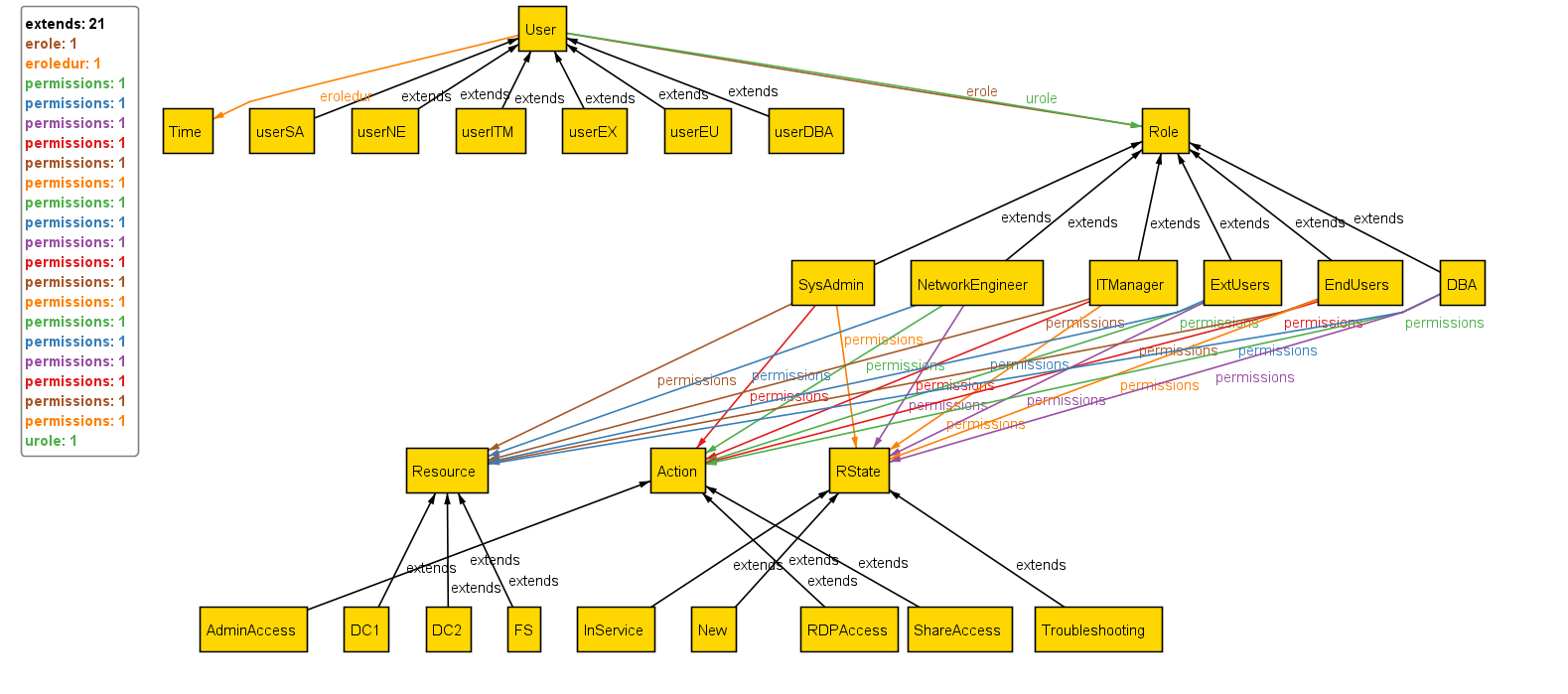}
	\caption{Policy 1 - Metamodel Output}
	\label{pol1_metamodel}
\end{figure}

\begin{figure}
	\centering
	\includegraphics[scale=.40]{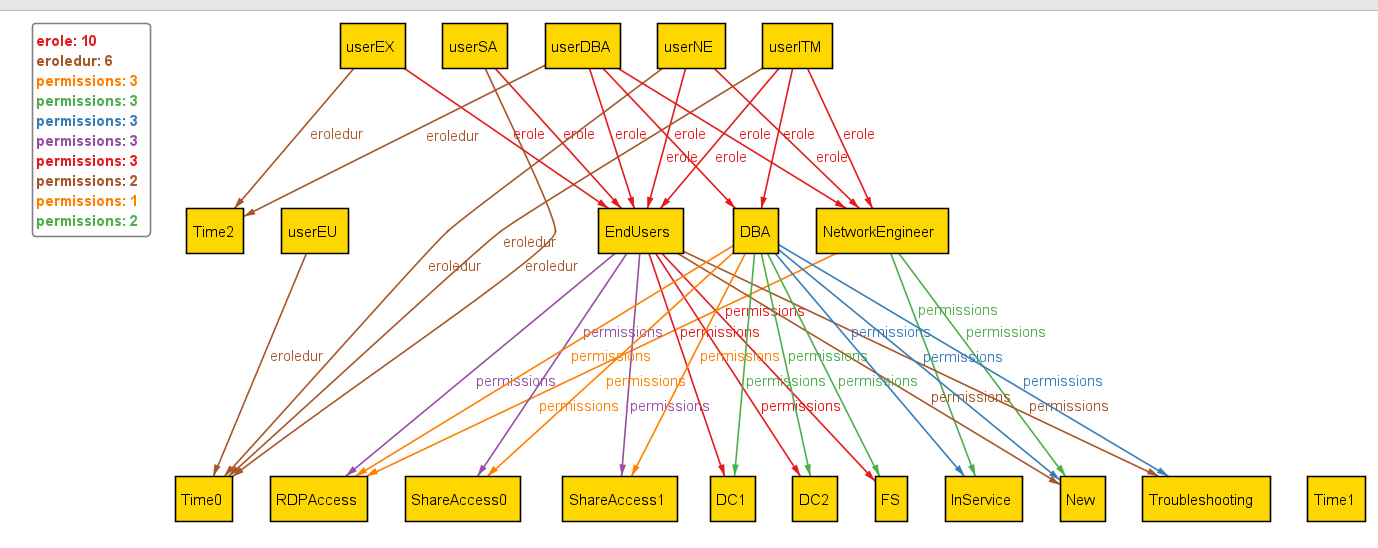}
	\caption{Policy 1 - Model Output}
	\label{pol1_alloy_model}
\end{figure}

With the predicates defined for assigning permissions, we added a trace function that defines different permissions for different states of the system. These permissions are assigned through Emergency role assignment and are valid only for a specific duration providing the temporal aspect of the model. Dedicated predicates for role assignment enforce SOD depending on the state.
\begin{lstlisting}[language=alloy]
//System Admin Role has all permissions 
pred trace
{
	init [T/first]
	let rs' = InService | roleassign[rs'] implies (exuPerms[rs'] and saPerms[rs'] and nePerms[rs'] and itmPerms[rs'] and dbaPerms[rs'] and euPerms[rs'])
	let rs'' = Troubleshooting | eroleassign[rs''] implies (exuPerms[rs''] and saPerms[rs''] and nePerms[rs''] and itmPerms[rs''] and dbaPerms[rs''] and euPerms[rs''])
}
\end{lstlisting}
With all the necessary components in place, the compilation was successful. Figure \ref{pol1_alloy_output} shows the compilation results. Figure \ref{pol1_metamodel} shows the metamodel generated by this.
\begin{table}[width=.9\linewidth,cols=2,pos=ht]
\caption{Policy 1 - Users}\label{pol1usersumtbl}
\begin{tabular*}{\tblwidth}{@{} LLLL@{} }
\toprule
User & Default Role\\
\midrule
userEx & External User\\
\midrule
userSA & System Administrator\\
\midrule
userNE & Network Engineer\\
\midrule
userDBA & Database Administrator\\
\midrule
userITM & IT Manager\\
\midrule
userEU & End-User\\
\bottomrule
\end{tabular*}
\end{table}
The model generated is shown in Figure \ref{pol1_alloy_model}
\subsection{Policy 2 - Alloy specification}
Based on the policy definition, below are the Sigs defined for Policy 2. This policy also has an additional object Environment. The resources are classified in to Production and Test based on the environment.
\begin{lstlisting}[language=alloy]
    abstract sig Role{} //defines the roles
    abstract sig Resource{} //defines the Resources
    abstract sig Action{} //defines the Actions or Access Levels
    abstract sig Env{} //defines the environment
\end{lstlisting}
Code to add time and state for adding the temporal aspects to the model.
\begin{lstlisting}[language=alloy]
    sig Time{} //for time definitions
    sig RState{}
\end{lstlisting}
With the definitions, several facts were added to model the policy. These facts enforce that every user must have one predefined role and the External Users role cannot be combined with any other roles. To demonstrate the role assignments, we created Users that are derived from the user Sig and will inherit the role(s). The roles are listed in Table \ref{pol1usersumtbl}. 
\begin{lstlisting}[language=alloy]
    fact {
    	no u:User | OtherUsers in u.urole and some (Developer+Tester+DBA+ServerAdmin+AppAdminUsers+AppUsers) & u.urole
    }
    fact {
    	all u:User | u.urole in Role
    }
\end{lstlisting}
Assertions are added to validate these facts. 
\begin{lstlisting}[language=alloy]
    assert rolesNeededforAll {
    	all u:User | u.urole in Role
    }
    assert OtherUserAccess{
    	no u:User | OtherUsers in u.urole and some (Developer+Tester+DBA+ServerAdmin+AppAdminUsers+AppUsers) & u.urole
    }
\end{lstlisting}
With the basic blocks defined, we added predicates to build the permissions. A predicate was defined for each role to define the permission set for that role that details all the actions that the role can perform on the resources. Eventually, we modified the code to include Environment as part of the permissions and rewrote all the permission predicates to include that. Below is one of the predicates.
\begin{lstlisting}[language=alloy]
    pred devPerms [rs:RState]
    {
    	//NO Access to Prod App, Prod and Test App Server, Prod and Test DB Server & Data
    	no ud:Developer, aps:AppSrvr, p:prod, t:test | ud.permissions = p + rs + aps + AdminAccess and ud.permissions = p + rs+ aps+ Access and ud.permissions = t + rs + aps + AdminAccess and ud.permissions = t + rs+ aps+ Access
    	no ud:Developer, dbs:DBSrvr, p:prod,t:test | ud.permissions = p + rs + dbs + AdminAccess and ud.permissions = p + rs+ dbs+ Access and ud.permissions = p + rs+ dbs+ DBAccess and ud.permissions = t + rs + dbs + AdminAccess and  ud.permissions = t + rs+ dbs+ Access and ud.permissions = t + rs+ dbs+ DBAccess
    	no ud:Developer, pd:ProdData | ud.permissions = rs+ pd + DataAccess
    	no ud:Developer, pa:App, p:prod | ud.permissions = p + rs + pa + AppAdminAccess and ud.permissions = p + rs+ pa + AppAccess
    	//Only Allowed access to Deid Data
    	all ud:Developer, td:DeidData | ud.permissions = rs+ td + DataAccess 
    	//Only Allowed access to Test App
    	all ud:Developer, ta:App, t:test | ud.permissions = t + rs + ta + AppAdminAccess and ud.permissions = t + rs+ ta + AppAccess	
    }
\end{lstlisting}

\begin{figure}
	\centering
	\includegraphics[scale=.50]{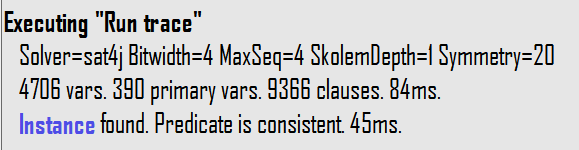}
	\caption{Policy 2 - Alloy Output}
	\label{pol2_alloy_output}
\end{figure}

\begin{figure}
	\centering
	\includegraphics[scale=.30]{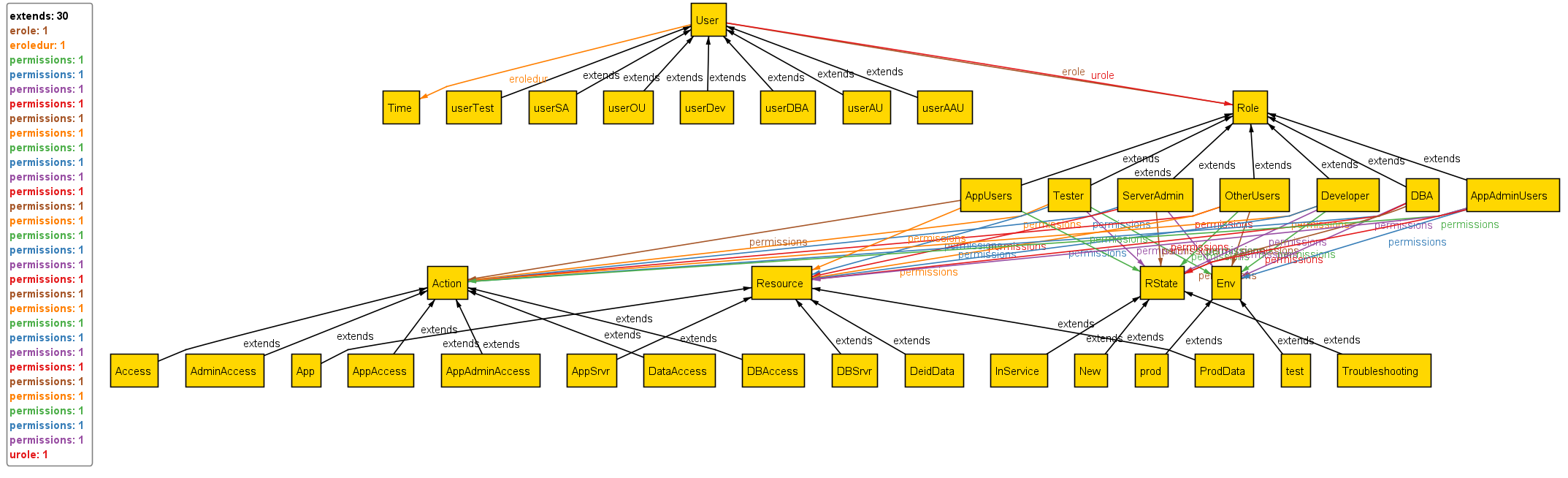}
	\caption{Policy 2 - Metamodel Output}
	\label{pol2_metamodel}
\end{figure}

\begin{figure}
	\centering
	\includegraphics[scale=.40]{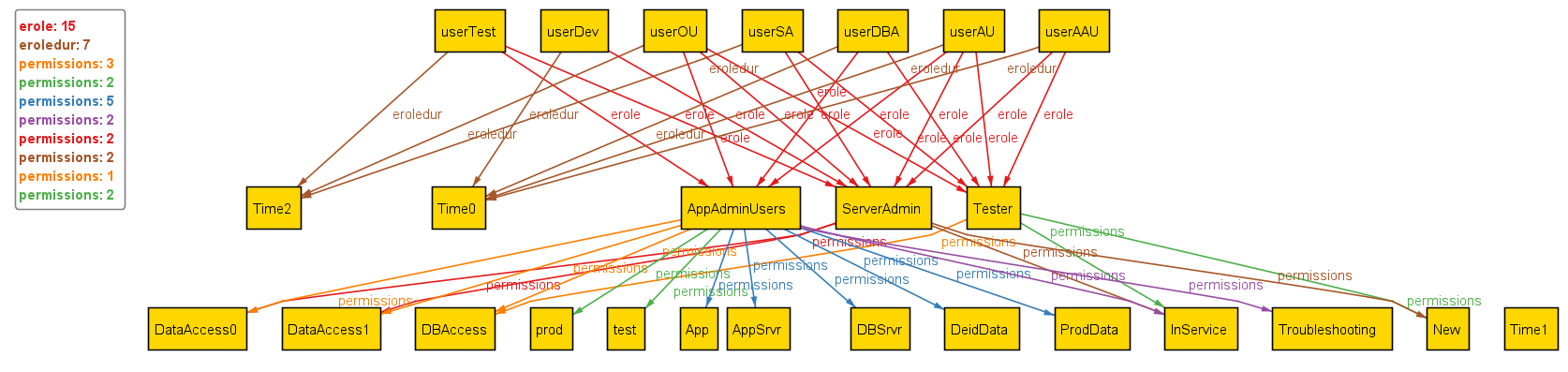}
	\caption{Policy 2 - Model Output}
	\label{pol2_alloy_model}
\end{figure}

With the predicates defined for assigning permissions, we added a trace function that defines different permissions for different states of the system. These permissions are assigned through Emergency role assignment and are valid only for a specific duration providing the temporal aspect of the model. Dedicated predicates for role assignment enforce SOD depending on the state.
\begin{lstlisting}[language=alloy]
    pred trace
    {
    	init [T/first]
    	let rs' = InService | roleassign[rs'] implies (ouPerms[rs'] and devPerms[rs'] and testPerms[rs'] and saPerms[rs'] and dbaPerms[rs'] and appadminPerms[rs'] and appuserPerms[rs'])
    	let rs' = Troubleshooting | eroleassign[rs'] implies (ouPerms[rs'] and devPerms[rs'] and testPerms[rs'] and saPerms[rs'] and dbaPerms[rs'] and appadminPerms[rs'] and appuserPerms[rs'])
    }
\end{lstlisting}
With all the necessary components in place, the compilation was successful. Figure \ref{pol2_alloy_output} shows the compilation results. Figure \ref{pol2_metamodel} shows the metamodel generated by this.
\begin{table}[width=.9\linewidth,cols=2,pos=ht]
\caption{Policy 2 - Users}\label{pol2usersumtbl}
\begin{tabular*}{\tblwidth}{@{} LLLL@{} }
\toprule
User & Default Role\\
\midrule
userDev & Developer\\
\midrule
userTest & Tester\\
\midrule
userSA & System Administrator\\
\midrule
userDBA & Database Administrator\\
\midrule
userAAU & Application Administrator\\
\midrule
userAU & Application End-User\\
\midrule
useOU & Other Users\\
\bottomrule
\end{tabular*}
\end{table}
The model generated is shown in Figure \ref{pol2_alloy_model}
\subsection{Policy 3 - Alloy specification}
Based on the policy definition, below are the Sigs defined for Policy 3. 
\begin{lstlisting}[language=alloy]
    abstract sig Role{} //defines the roles
    abstract sig Resource{} //defines the Resources
    abstract sig Action{} //defines the Actions or Access Levels
\end{lstlisting}
Code to add time and state for adding the temporal aspects to the model.
\begin{lstlisting}[language=alloy]
    sig Time{} //for time definitions
    sig RState{}
\end{lstlisting}
The facts added to the model are listed below. The roles are listed in Table \ref{pol3usersumtbl}. 
\begin{lstlisting}[language=alloy]
    //Fact 1 - All Users must have one of preDefined role
    fact {
    	all u:User | u.urole in Role
    }
    
    //Fact 2 - Users in this policy must have unique roles as this policy concerns log access 
    fact {
    	no u:User | ServerEngineer in u.urole and some (NetworkEngineer + SecurityEngineer) & u.urole
    }
\end{lstlisting}
Assertions are added to validate these facts. 
\begin{lstlisting}[language=alloy]
    //Assertion 1 - Each user must have one role
    assert rolesNeededforAll {
    	all u:User | u.urole in Role
    }
    //Assertion 3 - Users must have unique roles
    assert uniqueNWrole{
    	no u:User | NetworkEngineer in u.urole and some (ServerEngineer + SecurityEngineer) & u.urole
    }
\end{lstlisting}
Predicates were added to assign permissions to the roles
\begin{lstlisting}[language=alloy]
//Permissions for Network Engineer Role
pred nwPerms[rs:RState]
{
	//No Access to Servers and Logs
	no une:NetworkEngineer,dc:DC | une.permissions = rs + dc +AdminAccess and une.permissions =  rs + dc + LogReview 
	no une:NetworkEngineer,ap:AppSrvr | une.permissions = rs + ap +AdminAccess and une.permissions =  rs + ap + LogReview
	//Full Access to Network Devices and Logs
	all une:NetworkEngineer,fw:FireWall | une.permissions = rs + fw +AdminAccess and une.permissions =  rs + fw + LogReview
	all une:NetworkEngineer,sw:Switches | une.permissions = rs + sw +AdminAccess and une.permissions =  rs + sw + LogReview
	//No Access to Log Aggregator
	no une:NetworkEngineer,lg:LogAggregator | une.permissions = rs + lg +AdminAccess and une.permissions =  rs + lg + LogReview
}
\end{lstlisting}

\begin{figure}
	\centering
	\includegraphics[scale=.50]{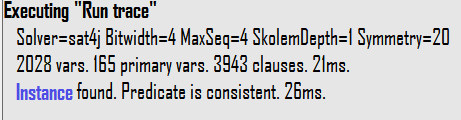}
	\caption{Policy 3 - Alloy Output}
	\label{pol3_alloy_output}
\end{figure}

\begin{figure}
	\centering
	\includegraphics[scale=.40]{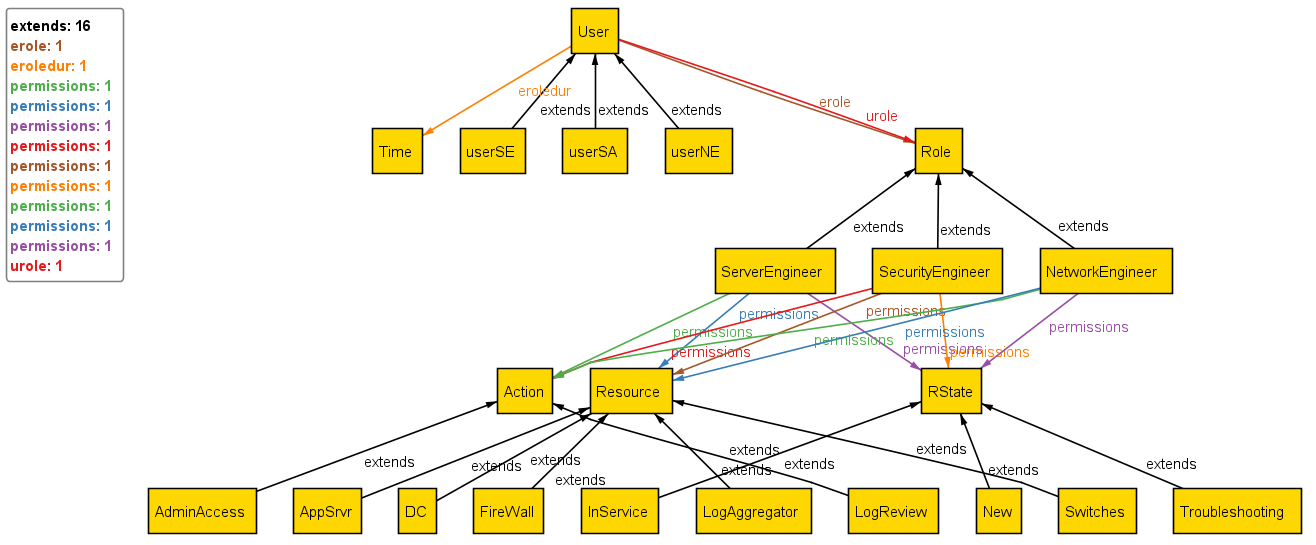}
	\caption{Policy 3 - Metamodel Output}
	\label{pol3_metamodel}
\end{figure}

\begin{figure}
	\centering
	\includegraphics[scale=.40]{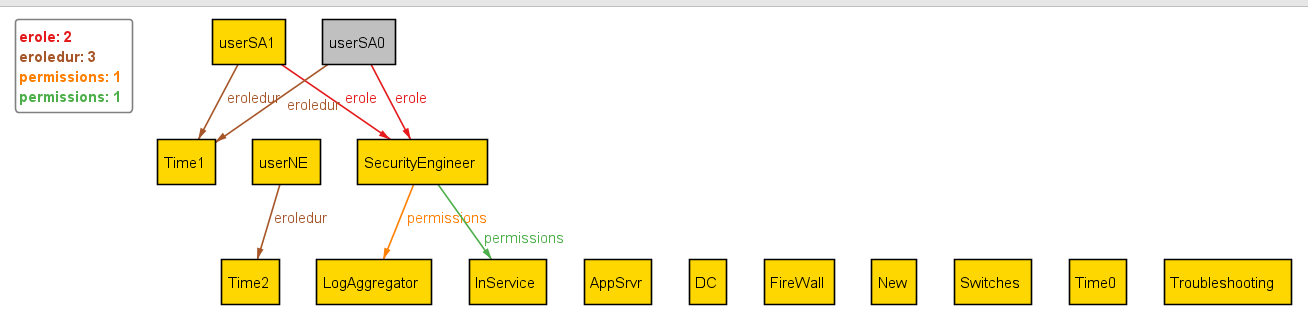}
	\caption{Policy 3 - Model Output}
	\label{pol3_alloy_model}
\end{figure}

With the predicates defined for assigning permissions, we added a trace function that defines different permissions for different states of the system. These permissions are assigned through Emergency role assignment and are valid only for a specific duration providing the temporal aspect of the model. 
\begin{lstlisting}[language=alloy]
    pred trace
    {
    	init [T/first]
    	let rs' = InService | roleassign[rs'] implies (sePerms[rs'] and saPerms[rs'] and nwPerms[rs'])
    	let rs' = Troubleshooting | eroleassign[rs'] implies (sePerms[rs'] and saPerms[rs'] and nwPerms[rs'])
    }
\end{lstlisting}
With all the necessary components in place, the compilation was successful. Figure \ref{pol3_alloy_output} shows the compilation results. Figure \ref{pol3_metamodel} shows the metamodel generated by this.
The model generated is shown in Figure \ref{pol3_alloy_model}
\begin{table}[width=.9\linewidth,cols=2,pos=ht]
\caption{Policy 3 - Users}\label{pol3usersumtbl}
\begin{tabular*}{\tblwidth}{@{} LLLL@{} }
\toprule
User & Default Role\\
\midrule
userSA & Server Engineer\\
\midrule
userSE & Security Engineer\\
\midrule
userNE & Network Engineer\\
\bottomrule
\end{tabular*}
\end{table}
\section*{Conclusion}
In this paper, a formal access control model is proposed for a healthcare system. We selected based users, roles and functions that can be application in most health sector domain as well as other relevant domains. We adopted state of the art RBAC access control model due to its ability to capture fine-grained access control requirement. We used Alloy Analyzer as a formal logic modeling tool to create and verify our model. We focused on some properties related to access control and showed using Alloy how such models can be implemented and also checked. Alloy includes also the ability to check for assertions to find possible ways to violate the model based on the model entities and constraints. 

\printcredits
\bibliographystyle{cas-model2-names}

\bibliography{cas-refs}


\bio{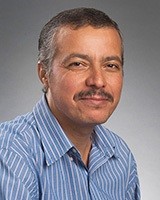}
Izzat Alsmadi is an Associate Professor in the department of computing and cyber security at the Texax A\&M, San Antonio. He has his master and PhD in Software Engineering from North Dakota State University in 2006 and 2008. He has more than 100 conference and journal publication. His research interests include: Cyber intelligence, Cyber security, Software security, software engineering, software testing, social networks and software defined networking. He is lead author, editor in several books including: Springer, The NICE Cyber Security Framework Cyber Security Intelligence and Analytics, 2019, Practical Information Security: A Competency-Based Education Course, 2018, Information Fusion for Cyber-Security Analytics (Studies in Computational Intelligence), 2016. The author is also a member of The National Initiative for Cybersecurity Education (NICE) group, which meets frequently to discuss enhancements on cyber security education at the national level.
\endbio

\bio{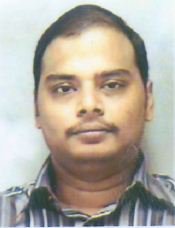}
I am currently pursuing my Masters in Cyber Security at Syracuse University. I received my bachelors degree in computer science from Annamalai Univeristy, Chidambaram, India. I started my professional career as System Administrator and held various roles including Developer and teacher. I hold two MCP certifications, MCSA certification as well as a WebSphere certification from IBM. In addition to my Masters, I am also working on advancing my MCSA certification.
\endbio

\end{document}